\documentclass[final]{l4dc2026}

\usepackage{mathtools}
\usepackage{comment}
\usepackage{siunitx}
\usepackage{subcaption} %
\usepackage{booktabs}   %
\usepackage{colortbl}   %
\usepackage{xcolor}     %
\newtheorem{assumption}{Assumption}
\usepackage[extramath,probability,reviewmark]{mylatexdefs}

\newcommand{\obstacle}[1]{obstacle\textendash\ensuremath{#1}\xspace}

\newcommand{\bb}[1]{\ensuremath{\mathbf{#1}}}
\newcommand{\bbsym}[1]{\ensuremath{\boldsymbol{#1}}}

\newcommand{\dd}{\mathsf{d}}

\title[A Hybrid L2O Framework for MIQP]{A Hybrid Learning-to-Optimize Framework for Mixed-Integer Quadratic Programming}
\usepackage{times}

\author{%
 \Name{Viet-Anh Le} \Email{vietanh@seas.upenn.edu}\\
 \Name{Mu Xie} \Email{mux2001@seas.upenn.edu}\\
 \Name{Rahul Mangharam} \Email{rahulm@seas.upenn.edu}\\
 \addr Department of Electrical \& Systems Engineering, University of Pennsylvania, Philadelphia, PA 19104, USA}
\begin{document}

\maketitle

\begin{abstract}%
In this paper, we propose a learning-to-optimize (L2O) framework to accelerate solving parametric mixed-integer quadratic programming (MIQP) problems, with a particular focus on mixed-integer model predictive control (MI-MPC) applications. 
The framework learns to predict integer solutions with enhanced optimality and feasibility by integrating supervised learning (for optimality), self-supervised learning (for feasibility), and a \emph{differentiable quadratic programming (QP) layer}, resulting in a hybrid L2O framework. 
Specifically, a neural network (NN) is used to learn the mapping from problem parameters to optimal integer solutions, while a differentiable QP layer is integrated to compute the corresponding continuous variables given the predicted integers and problem parameters. 
Moreover, a \emph{hybrid loss function} is proposed, which combines a supervised loss with respect to the global optimal solution, and a self-supervised loss derived from the problem's objective and constraints. 
The effectiveness of the proposed framework is demonstrated on two benchmark MI-MPC problems, with comparative results against purely supervised and self-supervised learning models.

\end{abstract}

\section{Introduction}

Mixed-integer optimization is fundamental to many control applications involving discrete decision-making, such as autonomous driving \citep{quirynen2024real}, traffic signal coordination with connected automated vehicles \citep{le2024distributed}, multi-robot pickup and delivery \citep{camisa2022multi}, motion planning and task assignment for robot fleets \citep{salvado2018motion}, and signal temporal logic specifications \citep{belta2019formal}. 
However, solving a mixed-integer program (MIP) is NP-hard because it requires combinatorial search over discrete decision variables. 
Consequently, computation time can grow exponentially with problem size or constraint complexity, making MIPs generally unsuitable for real-time control or decision-making.

Advancements in machine learning and differentiable programming provide a promising opportunity to accelerate MIP solvers through learning-to-optimize (L2O) frameworks. 
The literature on L2O for MIP problems is still limited but has gained increasing attention in recent years. 
The current state of the art can be categorized into two main approaches: (i) supervised learning (SL) and (ii) self-supervised learning (SSL). Classical SL approaches, \eg \citep{cauligi2021coco,cauligi2022prism,le2025multirobot}, train neural networks (NNs) to minimize the supervised loss between predictions and the optimal integer solutions generated by an optimization solver such as \texttt{GUROBI} \citep{gurobi}. 
A major drawback of SL approaches is that they do not guarantee feasibility, \ie the resulting convex continuous problems obtained by fixing the learned integer variables can be infeasible. SSL approaches \citep{tang2024learning,boldocky2025learning}, on the other hand, do not rely on labeled data and can improve feasibility by using a loss function that combines the objective function with a penalty for constraint violation. 
For example, \citep{tang2024learning} proposed a framework for mixed-integer nonlinear programming that solves the integer-relaxed problem, combined with integer correction layers to ensure integrality and a projection step to improve feasibility. \citep{boldocky2025learning} considered parametric MIQP problems within a differentiable predictive control framework, which constrains the integer solutions and optimal control inputs to a neural state-feedback law. 
However, trained SSL models may produce feasible but suboptimal solutions, since differentiable programming techniques such as gradient descent may converge to locally optimal solutions.

\begin{figure}[tb!]
\centering
\includegraphics[width=0.92\linewidth, trim=1.0cm 1.7cm 1.0cm 2.8cm, clip]{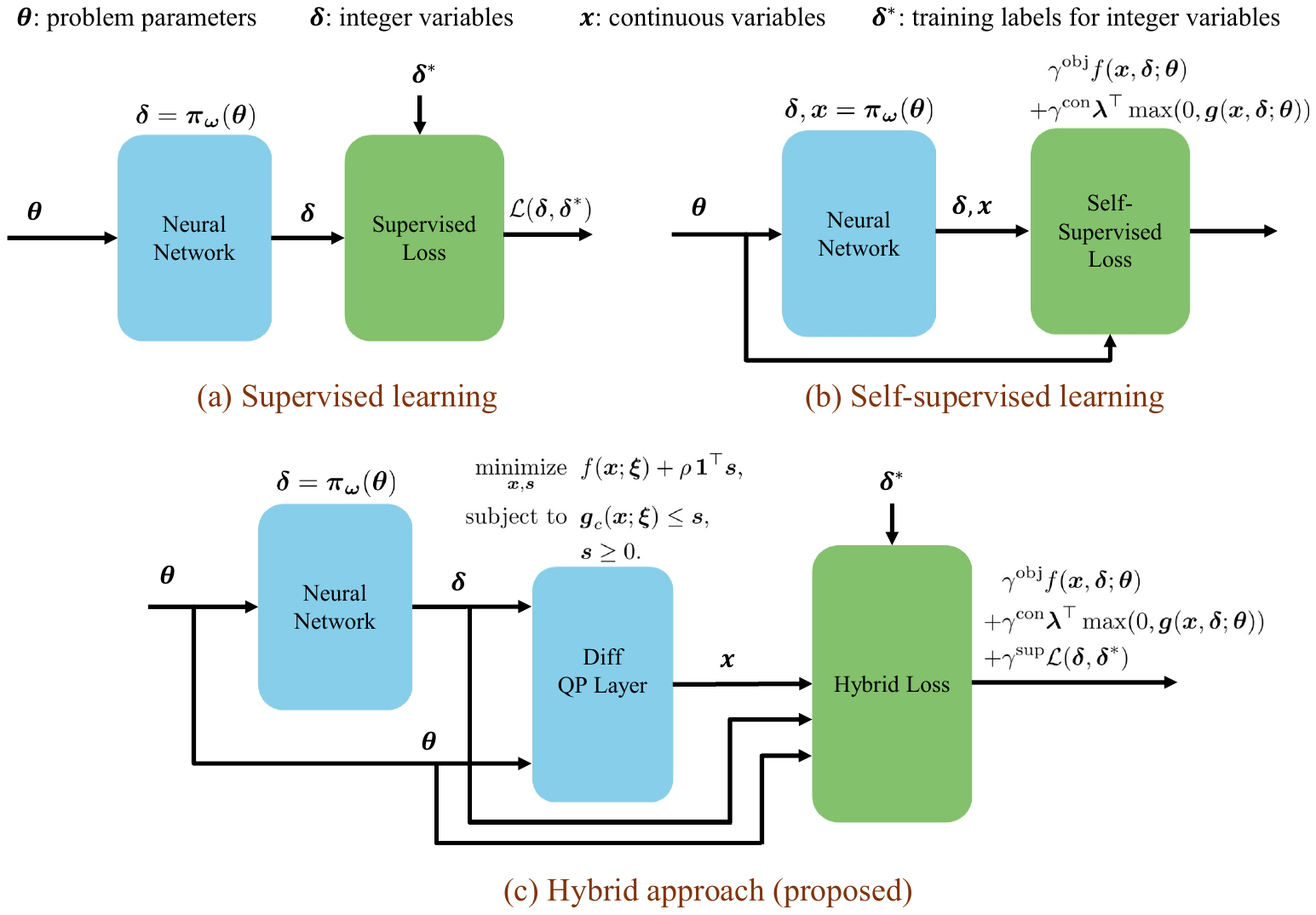}
\caption{Architecture of the proposed hybrid framework (c) compared with supervised learning (a) and self-supervised learning (b). 
In our framework, the NN takes the problem parameters $\bbsym{\theta}$ to predict the integer solution $\bbsym{\delta}$, while the QP layer computes the continuous solution $\bbsym{x}$ based on $\bbsym{\theta}$ and $\bbsym{\delta}$. 
In conventional SL and SSL, the NN is trained to predict the integer solution without considering the continuous solution or to predict both the integer and continuous solutions, respectively.}
\label{fig:model}
\vspace{-5mm}
\end{figure}

Addressing the limitations of both SL (infeasibility) and SSL (suboptimality), we propose a novel hybrid L2O framework that strategically combines both training paradigms with an integrated differentiable QP layer. 
The hybrid approach proposed in this paper shares conceptual similarities with physics-informed machine learning (PINN) by embedding the known optimization structure of the problem (the QP layer) as an inductive bias, much as PINNs embed physical laws (the PDEs).
First, we propose a novel architecture in which a differentiable QP layer is integrated into the network to better incorporate the optimization structure. 
During training, the QP layer takes the NN predictions of the integer decision variables as input and outputs the corresponding optimal solutions for the continuous variables, but it may be infeasible.
To address this issue and ensure that gradients can always be computed, we propose a simple yet effective approach that introduces a differentiable layer for the relaxed QP problem.
We prove that, if the penalty weight is chosen sufficiently large, the relaxed problem yields either the exact optimal solution when the original QP is feasible or the minimally infeasible solution otherwise. 
Second, we propose a new loss function design, called a \emph{hybrid loss function}, defined as a weighted sum of SL and SSL losses. 
This approach allows the framework to balance the feasibility-optimality trade-off in L2O.
The overall architecture of our framework in comparison with SL and SSL can be illustrated in Fig.~\ref{fig:model}.

Our framework differs from existing work in the relevant literature in the following aspects. First, a major difference lies in how we encode the dependency between continuous variables, integer variables, and problem parameters during training. 
In SL approaches \citep{cauligi2021coco,cauligi2022prism}, the goal is to learn the mapping from problem parameters to the optimal integer variables, while disregarding the continuous variables during training. 
In online prediction, the continuous variables are then obtained by solving a QP given the NN predictions of the integer variables. 
In contrast, SSL approaches \citep{tang2024learning,boldocky2025learning} train NNs to predict both the continuous and integer variables from the problem parameters. 
Thus, the prediction obtained directly from the NN does not explicitly account for the dependency between continuous and integer variables. 
In our approach, we incorporate the continuous variables into training by integrating a QP layer, based on the fact that the optimal continuous variables are the solutions of a parametric QP given the integer variables and the MIQP problem parameters. 
Therefore, our approach better incorporates the underlying optimization structure into both training and prediction than supervised and self-supervised learning.
Second, we combine supervised and self-supervised learning objectives to define a hybrid loss function, rather than relying on either one alone. 
This hybrid loss allows the framework to balance the optimality of supervised learning with training labels and the feasibility improvement of self-supervision during training. Thus, our proposed framework can be viewed as a compromise between SL and SSL. 
We show through numerical examples that the hybrid loss function achieves near-global optimality and minimal constraint violation for most problem instances.

\section{Preliminaries}

This section provides a brief discussion on the parametric MIQPs and L2O for accelerating solving MIQP problems.

\subsection{Parametric MIQPs}

We consider a parametric MIQP problem that takes the following form:
\begin{subequations}
\label{eq:compact-prb}
\begin{align}
\underset{\bbsym{x} \in \XXX, \bbsym{\delta} \in \III}{\minimize} & \;\; f (\bbsym{x}, \bbsym{\delta}; \bbsym{\theta}), \label{eq:compact-prb-obj} \\
\text{subject to} & \;\;  \bbsym{g} (\bbsym{x}, \bbsym{\delta}; \bbsym{\theta}) \le 0, \end{align}
\end{subequations}
where $\bbsym{x}$ is the vector of continuous optimization variables, $\bbsym{\delta}$ is the vector of integer optimization variables, and $\bbsym{\theta}$ is the vector of problem parameters.
We let $\XXX$ and $\III$ be the domains for continuous and integer optimization variables, and $\bbsym{g} (\cdot) = [g_1 (\cdot), \dots, g_r (\cdot)]$ be the vector of $r$ linear constraints.
We assume that $\XXX$ and $\III$ are non-empty.
In this work, we consider a convex quadratic objective function and linear constraints, \ie 
\begin{equation}
\begin{split}
f(\bbsym{x}, \bbsym{\delta}; \bbsym{\theta}) &= \frac{1}{2} \begin{bmatrix} \bbsym{x}\\ \bbsym{\delta} \end{bmatrix}^\top \bbsym{Q}(\bbsym{\theta}) \begin{bmatrix} \bbsym{x}\\ \bbsym{\delta} \end{bmatrix} + \bbsym{p}(\bbsym{\theta})^\top \begin{bmatrix} \bbsym{x}\\ \bbsym{\delta} \end{bmatrix}, \\
\bbsym{g} (\bbsym{x}, \bbsym{\delta}; \bbsym{\theta}) &= \bbsym{G}(\bbsym{\theta}) \begin{bmatrix} \bbsym{x}\\ \bbsym{\delta} \end{bmatrix} - \bbsym{h} (\bbsym{\theta}).
\end{split}
\end{equation}
where $\bbsym{Q}(\bbsym{\theta}) \succeq 0$.
Note that given known parameters $\bbsym{\theta}$ and integer variables $\bbsym{\delta}$, the optimal solution of the continuous variables can be obtained by solving a QP problem, if it is feasible, as follows:
\begin{subequations}
\label{eq:qp}
\begin{align}
\underset{\bbsym{x} \in \XXX}{\minimize} & \;\; f (\bbsym{x}; \bbsym{\xi}), \\
\text{subject to} & \;\; \bbsym{g}_c (\bbsym{x}; \bbsym{\xi}) \le 0,
\end{align}
\end{subequations}
where $\bbsym{g}_c (\cdot)$ denotes the components of $\bbsym{g} (\cdot)$ that involve at least one continuous decision variable, and $\bbsym{\xi} = [\bbsym{\delta}^\top, \bbsym{\theta}^\top]^\top$.
The objective function and constraint function in \eqref{eq:qp} can be expressed as:
\begin{align}
f (\bbsym{x}; \bbsym{\xi}) &= \frac{1}{2} \bbsym{x}^\top \bbsym{Q}_{x}(\bbsym{\xi}) \bbsym{x} + \bbsym{p}_{x}(\bbsym{\xi})^\top  \bbsym{x}, \\
\bbsym{g}_c (\bbsym{x}; \bbsym{\xi}) &= \bbsym{G}_{x}(\bbsym{\xi}) \bbsym{x} - \bbsym{h}_{x}(\bbsym{\xi}).    
\end{align}

\textbf{Mixed-integer MPC:} A typical application that can greatly benefit from L2O approaches is model predictive control (MPC).
In MPC, we solve a parametric optimization problem at every time step, in which the problem parameters may include, for instance, a given initial state, target state, to name a few.
Using machine learning to solve or assist in solving parametric MPC problems enables real-time implementation of complex MPC, such as nonlinear MPC or mixed-integer MPC.
We consider the state $\bbsym{x}_t \in \XXX \subset \mathbb{R}^{n_x}$ and control inputs $\bbsym{u}_t \in \UUU \subset \mathbb{R}^{n_u}$ as the continuous decision variables and let $\bbsym{\delta}_t \in \III \subset \ZZ^{n_\delta}$ be the integer decision variables. 
Let $H$ be the control horizon length, and $\bbsym{x}_{0:H}, \bbsym{u}_{0:H-1}, \bbsym{\delta}_{0:H-1}$ be the concatenated vectors over the control horizon.
Given a vector of problem parameters $\bbsym{\theta} \in \mathbb{R}^{n_p}$, a parametric MI-MPC can be written as:
\begin{equation}
\begin{aligned}
\minimize_{\bbsym{x}_{0:H}, \bbsym{u}_{0:H-1}, \bbsym{\delta}_{0:H-1}} \; & c_H(\bbsym{x}_H; \bbsym{\theta}) + \sum_{t=0}^{H-1}\; c_t(\bbsym{x}_t, \bbsym{u}_t, \bbsym{\delta}_t; \bbsym{\theta}), \\
\text{subject to \quad} \; 
& \bbsym{x}_0 = \bbsym{x}_{\text{init}}(\bbsym{\theta}), \\
& \bbsym{x}_{t+1} = \bbsym{f} (\bbsym{x}_t, \bbsym{u}_t, \bbsym{\delta}_t; \bbsym{\theta}), \; t = 0, \ldots, H-1, \\
& \bbsym{g}_{t}(\bbsym{x}_t, \bbsym{u}_t, \bbsym{\delta}_t; \bbsym{\theta}) \le 0,\; t = 0, \ldots, H-1, \\
& \bbsym{g}_{H} (\bbsym{x}_t; \bbsym{\theta}),
\end{aligned}
\label{eq:miocp}
\end{equation}
where the stage cost $c_t(\cdot)$ and terminal cost $c_H (\cdot)$ are convex quadratic, while the dynamics $\bbsym{f}(\cdot)$, the inequality constraints $\bbsym{g}_{t}(\cdot)$ and $\bbsym{g}_{H}(\cdot)$ are assumed to be linear functions. 
The objective function and constraints are functions of the parameter vector $\bbsym{\theta} \in \Theta$, where $\Theta \subseteq \mathbb{R}^{n_p}$ is the admissible set of parameters.

\subsection{Learning to Optimize for MIQPs}

Due to the combinatorial nature, the problem \eqref{eq:compact-prb} is computationally demanding to solve, where finding the optimal solution may scale exponentially with the problem size.
Meanwhile, solving \eqref{eq:qp} while the integers are fixed is significantly cheaper to solve than the MIQP.
An interesting approach to accelerating solving problems of the form \eqref{eq:compact-prb} is to learn a map between the vector of problem parameters $\bbsym{\theta}$ and the discrete optimizer $\bbsym{\delta}^*$ by an NN, $\bbsym{\delta}^* = \bbsym{\pi}_{\bbsym{\omega}}(\bbsym{\theta})$, where $\bbsym{\omega}$ is the vector of network weights.
Overall, there are two main approaches for learning $\bbsym{\pi}_{\bbsym{\omega}}$, including (i) supervised learning and (ii) self-supervised learning.

\textbf{Supervised Learning:} 
The NN $\bbsym{\pi}_{\bbsym{\omega}}$ can be trained by a classical SL approach. 
In SL, we collect the optimal solutions $\bbsym{\delta}^{i,*}$ corresponding to each $\bbsym{\theta}^i$, for $i = 1, \dots, M$, obtained from a solver.
We then use the dataset $\{\bbsym{\theta}^i, \bbsym{\delta}^{i,*}\}$ to train a NN that minimizes the following loss function:
\begin{equation}
\minimize_{\bbsym{\omega}} \; \frac{1}{M} \; \sum_{i=1}^M \; \LLL (\bbsym{\pi}_{\bbsym{\omega}}(\bbsym{\theta}^i), \bbsym{\delta}^{i,*}),
\end{equation}
where $\LLL$ denotes a supervised loss (\eg Huber or cross-entropy) between the predicted outputs and labels.
However, the prediction from an SL model may not ensure that the resulting QP is feasible, although the original MIQP is feasible.

\textbf{Self-Supervised Learning:}
Self-supervised learning, contrary to SL, does not rely on labeled data for training the model.
It instead trains models by minimizing the objective function and constraint violation directly from model predictions.
In generic SSL, an NN is trained to predict both the integer and continuous variables, \ie  $(\bbsym{\delta}^i, \bbsym{x}^i) = \bbsym{\pi}_{\bbsym{\omega}}(\bbsym{\theta}^i)$,
and to train the NNs given a dataset with $M$ training instances, the following self-supervised loss function is used:
\begin{subequations}
\label{eq:loss-fcn}
\begin{align}
\underset{\bbsym{\omega}}{\minimize} & \;\; \frac{1}{M} \sum_{i=1}^{M} f (\bbsym{x}^i, \bbsym{\delta}^i; \bbsym{\theta}^i) + \bbsym{\lambda}^\top \max \big(0, \bbsym{g} (\bbsym{x}^i, \bbsym{\delta}^i; \bbsym{\theta}^i) \big), \label{eq:loss} \\
\text{subject to} & \;\; (\bbsym{\delta}^i, \bbsym{x}^i) = \bbsym{\pi}_{\bbsym{\omega}}(\bbsym{\theta}^i),
\end{align}
\end{subequations}
In \eqref{eq:loss-fcn}, we include a penalty for constraint violation with max penalty (implemented via a ReLU) function.
$\bbsym{\lambda} \in \RR_{>0}^r$ is a vector of penalty parameters that balances the trade-off between minimizing the objective function and satisfying the constraints.
Although the penalty methods lack formal guarantees, they often outperform their hard constraint counterparts in practice.
Training the NN with a self-supervised loss function can improve, though it does not guarantee, feasibility.
Nevertheless, a major drawback of this approach is that since the self-supervised loss \eqref{eq:loss-fcn} is non-convex with respect to $\bbsym{\omega}$, gradient-based methods may not converge and may converge to a sub-optimal solution.
Moreover, in MPC applications, designing an NN to predict the optimal continuous decision variables from the problem parameters is generally challenging, as it is difficult to enforce the system dynamics on the network outputs \citep{cauligi2021coco}, unless the optimal integer and continuous solutions at each time step are constrained to follow a state-feedback law, as in differentiable predictive control \citep{boldocky2025learning}.

\begin{remark}
\textbf{(Differentiating through discrete operations)} 
In many approaches, the NNs are designed to directly output discrete values, using discrete operations such as rounding to produce those outputs. 
These discrete operations lead to non-differentiability and hinder the use of standard differentiable programming for network training. A common approach to address this issue is the straight-through estimator (STE) \citep{bengio2013estimating}, which enables backpropagation through discrete operations.
During the forward pass, STE applies a non-differentiable operation to obtain discrete values. 
During the backward pass, it bypasses the non-existent gradients of these operations by replacing them with those of smooth surrogate functions.
This approach was used in self-supervised learning frameworks for mixed-integer programming \citep{tang2024learning,boldocky2025learning}.
We also use this technique in our framework.
\end{remark}

We observe that the strength of SL in finding global solutions corresponds to the weakness of SSL, and vice versa, the strength of SSL in improving feasibility corresponds to the weakness of SL.
Therefore, an interesting idea is to combine SL and SSL to exploit the benefits of both approaches.

\section{Proposed Framework}

In this section, we present an L2O framework for MIQPs in which a differentiable QP layer is incorporated, and a hybrid loss function combining SL and SSL is proposed.

\subsection{Differentiable QP Layers for Feasible and Infeasible Problems}

Given known $\bbsym{\delta}^*$ and $\bbsym{\theta}$, the optimal solution of the continuous variables can be obtained by solving the QP problem \eqref{eq:qp}, if it is feasible.
Thus, we can consider the QP problem as a differentiable layer within deep learning architectures, denoted by $\bbsym{x} = \mathrm{QP} (\bbsym{\delta}, \bbsym{\theta})$.
Therefore, it leads to the following problem in which we approximate the integer solutions by NNs, and the continuous solutions by a QP layer: 
\begin{subequations}
\label{eq:pen-prb}
\begin{align}
\underset{\bbsym{\omega}}{\minimize} & \;\; f (\bbsym{x}, \bbsym{\delta}; \bbsym{\theta}) , \\
\text{subject to} & \;\; \bbsym{\delta} = \bbsym{\pi}_{\bbsym{\omega}}(\bbsym{\theta}), \\
& \;\; \bbsym{x} = \mathrm{QP} (\bbsym{\delta}, \bbsym{\theta}), \\
& \;\; \bbsym{g} (\bbsym{x}, \bbsym{\delta}; \bbsym{\theta}) \le 0.
\end{align}
\end{subequations}
To ensure the validity of the L2O framework using a QP layer and differentiable programming, we need the following assumption.

\begin{assumption}
\label{assp:convexity}
The QP problem \eqref{eq:qp} is strictly convex.
\end{assumption}
As stated in \citet[Theorem 1]{amos2017optnet}, Assumption~\ref{assp:convexity} is needed to ensure that the QP layer is subdifferentiable everywhere, and differentiable at all but a measure-zero set of points, because the solution of a strictly convex QP is continuous.
The question is how to compute the gradient of the optimal solution $\bbsym{x}^*$ with respect to the argument, \ie $\frac{\partial \bbsym{x}^*}{\partial \bbsym{\xi}}$.
If the problem is feasible, these derivatives can be obtained by differentiating the KKT conditions (sufficient and necessary conditions for optimality) of $\eqref{eq:qp}$.
However, since the methods in \citep{amos2017optnet,agrawal2019differentiable} rely on KKT conditions, it assumes the QP problem is feasible.
On the other hand, in our framework, the optimization problems might be infeasible during training, given different values of the integer variables from the NN.
Thus, we cannot directly incorporate the differentiable QP layer in \citep{amos2017optnet,agrawal2019differentiable} into our framework.
In this section, we present a simple yet efficient way to handle infeasibility during training, as described below. 

To this end, we introduce slack variables $\bbsym{s}$ for the constraints, leading to the following QP:
\begin{subequations}
\label{eq:qp-slack}
\begin{align}
\underset{\bbsym{x} \in \XXX, \bbsym{s}}{\minimize} & \;\; f (\bbsym{x}; \bbsym{\xi}) + \rho\, \bbsym{1}^\top \bbsym{s}, \\
\text{subject to} & \;\; \bbsym{g}_c (\bbsym{x}; \bbsym{\xi}) \le \bbsym{s}, \\
& \;\; \bbsym{s} \ge 0.
\end{align}
\end{subequations}
For ease of notations, in the rest of this section, we omit the argument $\bbsym{\xi}$ while mentioning the terms involving it.
Since \eqref{eq:qp-slack} is feasible given any realization of $\bbsym{\xi}$ as long as the domain for $\bbsym{x}$ is non-empty, we can apply the KKT conditions.
First, we formulate the Lagrangian of \eqref{eq:qp-slack} as follows:
\begin{equation}
L(\bbsym{x}, \bbsym{s}, \bbsym{\mu}, \bbsym{\kappa}) = 
\frac{1}{2} \bbsym{x}^\top \bbsym{Q}_{x} \bbsym{x} + \bbsym{p}_{x}^\top  \bbsym{x} 
+ \rho\, \bbsym{1}^\top \bbsym{s}
+ \bbsym{\mu}^\top (\bbsym{G}_{x} \bbsym{x} - \bbsym{h}_{x} - \bbsym{s}) - \bbsym{\kappa}^\top \bbsym{s},
\end{equation}
where $\bbsym{\mu} \geq 0$ and $\bbsym{\kappa} \geq 0$ are the dual variables on the constraints, and $\rho > 0$ is a penalty weight for the slack variables.
The KKT conditions for stationarity, primal feasibility, and complementary slackness are
\begin{subequations}
\label{eq:kkt}
\begin{align}
\bbsym{Q}_{x} \bbsym{x}^* + \bbsym{p}_{x} 
+ \bbsym{G}_{x}^\top \bbsym{\mu}^* &= 0, \\
\rho\, \bbsym{1} - \bbsym{\mu}^* - \bbsym{\kappa}^* &= 0, \label{eq:kkt-b} \\ 
D (\bbsym{\mu}^*) \big(\bbsym{G}_{x} \bbsym{x}^* - \bbsym{h}_{x} - \bbsym{s}^*\big) &= 0, \\
D (\bbsym{\kappa}^*) \bbsym{s}^* &= 0,
\end{align}
\end{subequations}
where $D(\cdot)$ is the operation creating a diagonal matrix from a vector.
Taking the differentials of the KKT conditions \eqref{eq:kkt}, we obtain
\begin{subequations}
\begin{align}
\dd \bbsym{Q}_{x} \bbsym{x}^* + \bbsym{Q}_{x} \dd \bbsym{x} + \dd \bbsym{p}_{x} 
+ \dd \bbsym{G}_{x}^\top \bbsym{\mu}^* + \bbsym{G}_{x}^\top \dd \bbsym{\mu} & = 0, \\
\dd \bbsym{\mu} + \dd \bbsym{\kappa} & = 0, \\
D\big(\bbsym{G}_{x} \bbsym{x}^* - \bbsym{h}_{x} - \bbsym{s}^*\big) \dd \bbsym{\mu} + D(\bbsym{\mu}^*)(\dd \bbsym{G}_{x} \bbsym{x}^* + \bbsym{G}_{x} \dd \bbsym{x} - \dd \bbsym{h}_{x} - \dd \bbsym{s}) &= 0, \\
D(\bbsym{s}^*) \dd \bbsym{\kappa} + D (\bbsym{\kappa}^*) \dd \bbsym{s} &= 0,
\end{align}
\end{subequations}
or in the matrix form as follows:
\begin{equation}
\hspace{-1mm}
\small{
\begin{multlined}    
\begin{bmatrix}
\bbsym{Q}_x & \bbsym{0} & \bbsym{G}_x &  \bbsym{0} \\
D(\bbsym{\mu}^*) \bbsym{G}_{x} & -D(\bbsym{\mu}^*) & D\big(\bbsym{G}_{x} \bbsym{x}^* - \bbsym{h}_{x} - \bbsym{s}^*\big) & \bbsym{0} \\
\bbsym{0} & D (\bbsym{\kappa}^*) & \bbsym{0} & D(\bbsym{s}^*)  \\
\bbsym{0} & \bbsym{0} & \bbsym{I} & \bbsym{I} \\
\end{bmatrix} 
\begin{bmatrix}
\dd \bbsym{x} \\
\dd \bbsym{s} \\
\dd \bbsym{\mu} \\
\dd \bbsym{\kappa} \\
\end{bmatrix} 
=
\begin{bmatrix}
- \dd \bbsym{Q}_{x} \bbsym{x}^* - \dd \bbsym{p}_{x} 
- \dd \bbsym{G}_{x}^\top \bbsym{\mu}^* \\
- D(\bbsym{\mu}^*)(\dd \bbsym{G}_{x} \bbsym{x}^* - \dd \bbsym{h}_{x}) \\
\bbsym{0} \\
\bbsym{0} \\
\end{bmatrix}.
\end{multlined}
}
\label{eq:kkt-diff}
\end{equation}
Using these equations, we can form the Jacobians of $\bbsym{x}^*$ and $\bbsym{s}^*$ with respect to any of the problem parameters.
The details can be found in \citep{amos2017optnet}, \citep{agrawal2019differentiable}.

Note that given any $\bbsym{x}$, $\bbsym{s}^* = \max (0, \bbsym{g}_c (\bbsym{x}))$.
In other words, the use of slack variables in \eqref{eq:qp-slack} is equivalent to using the max penalty function.
The following theorem shows that if the original QP \eqref{eq:qp} is feasible, then by choosing a sufficiently large value for $\rho$, solving \eqref{eq:qp-slack} yields the same solution as \eqref{eq:qp}.
\begin{theorem}
If the original QP \eqref{eq:qp} is feasible and let $\bbsym{x}^{'*}$ and $\bbsym{\mu}^{'*}$ be the optimal solutions and multipliers of the problem, respectively. 
If the penalty weight $\rho$ is chosen such that $\rho> \norm{\bbsym{\mu}^{'*}}_{\infty}$,
then the optimal solutions of \eqref{eq:qp-slack} and the original QP \eqref{eq:qp} are the same.
\end{theorem}

The proof of this theorem follows directly from Proposition 5.25 in \citep{bertsekas2014constrained} and is therefore omitted.
Therefore, if the original QP is feasible, we get the exact derivative of the optimal solution by using the KKT conditions of the relaxed problem. 
In the infeasible case, we show in the following theorem that, under some mild conditions, the obtained solution from \eqref{eq:qp-slack} is a point that minimizes the constraint violation, and among all the solutions with minimal constraint violation, the obtained solution also minimizes the original objective function.

\begin{theorem}
\label{thrm:inf}
If the original QP \eqref{eq:qp} is infeasible and assume that either one of the following properties hold:
\begin{itemize}
\item $\bbsym{Q}_x \succ 0$, which means $f(\bbsym{x})$ is coercive.
\item $\XXX$ is compact (closed and bounded).
\end{itemize}
Let us denote $v(\bbsym{x}) := \bbsym{1}^\top \max (0, \bbsym{g}_c (\bbsym{x}))$.
If the penalty weight $\rho$ is chosen sufficiently large,
then (i) a minimizer $(\bbsym{x}^*_\rho, \bbsym{s}^*_\rho)$ of \eqref{eq:qp-slack} achieves minimal total violation, \ie $v (\bbsym{x}^*_\rho) = v^*$ and (ii)
\begin{equation}
\label{eq:x-rho-star}
\bbsym{x}^*_\rho \in \underset{\bbsym{x} \in \XXX}{\argmin} \; \{f(\bbsym{x}) \,:\, v(\bbsym{x}) = v^* \}.    
\end{equation}
\end{theorem}

The proof for Theorem~\ref{thrm:inf} is given in Appendix~\ref{apx:proof}.

\subsection{Hybrid Training Loss}

Our training framework relies on a \emph{hybrid loss function} that combines supervised and self-supervised losses.
The main advantage of this hybrid loss is that it leverages the strengths of supervised learning (SL) in achieving global solution optimality and self-supervised learning (SSL) in improving constraint satisfaction.
The hybrid loss used to train the neural network can be defined as follows:
\begin{equation}
\underset{\bbsym{\omega}}{\minimize} \;\; \frac{1}{M} \sum_{i=1}^{M} \gamma^{\mathrm{obj}} f (\bbsym{x}^i, \bbsym{\delta}^i; \bbsym{\theta}^i) 
+ \gamma^{\mathrm{con}} \bbsym{\lambda}^\top \max \big(0, \bbsym{g} (\bbsym{x}^i, \bbsym{\delta}^i; \bbsym{\theta}^i) \big) 
+ \gamma^{\mathrm{sup}} \LLL (\bbsym{\delta}^i, \bbsym{\delta}^{i,*}),
\label{eq:hybrid-loss}     
\end{equation}
where $\gamma^{\mathrm{obj}}$, $\gamma^{\mathrm{con}}$, and $\gamma^{\mathrm{sup}} \in \RR_{\ge 0}$ are the weights for the objective value, constraint violation, and supervised loss, respectively.
Note that in our framework, the constraints involving at least one continuous variable can be directly handled using the differentiable QP layer, while the constraints involving only the integer variables must be incorporated into the loss function.

\section{Results and Discussions}

We validate our hybrid L2O framework on two benchmark MI-MPC problems: (i) collision avoidance for robot navigation and (ii) simplified thermal energy tanks \citep{boldocky2025learning}.
In the first example, binary variables are used to formulate the collision-avoidance constraints, thus, there are several coupling constraints between the integer and continuous variables.
Meanwhile, the second example involves integer decision variables and demonstrates the case where the integers appear in the objective function.
The details of the two examples are provided in Appendix~\ref{apx:example}.
For each example, a multilayer perceptron network is constructed with four hidden layers, 128 neurons per layer, and ReLU activation functions.
Our implementation and examples are available at \url{https://github.com/mlab-upenn/L2O-MIQP}.

\begin{figure}[tb!]
\centering
\includegraphics[width=0.95\textwidth]{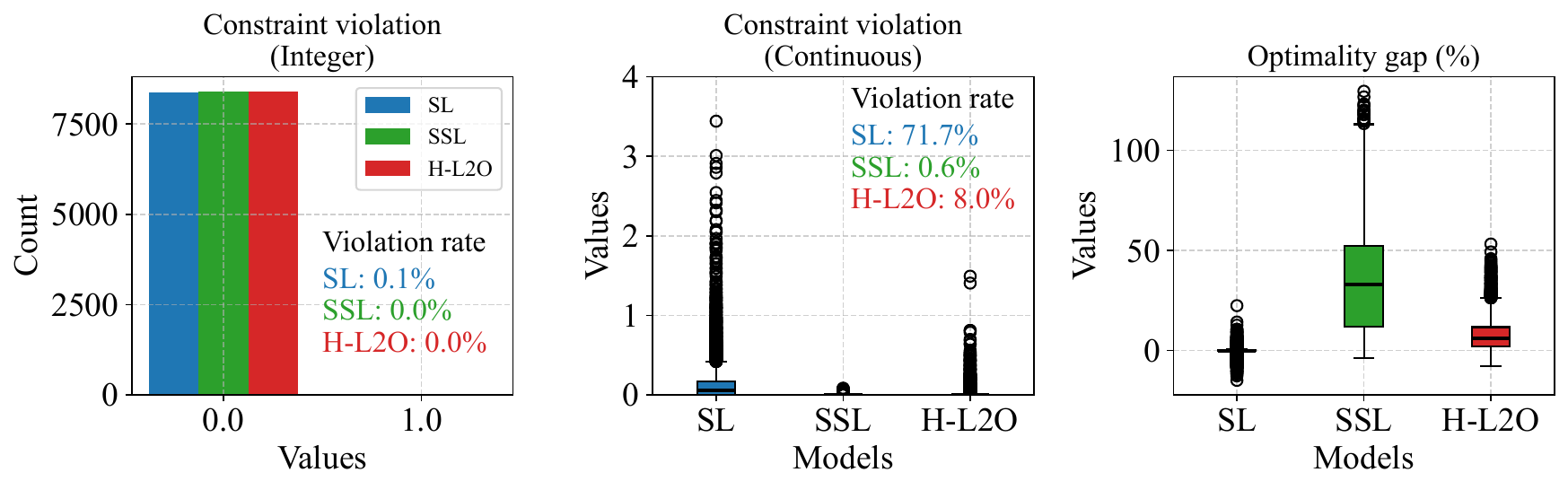}
\caption{Statistical comparison of the three models: hybrid L2O (H-L2O), supervised learning (SL), and self-supervised learning (SSL), for the robot navigation example.}
\label{fig:robot_nav}
\vspace{-5mm}
\end{figure}

\begin{figure}[tb!]
\centering
\includegraphics[width=0.95\textwidth]{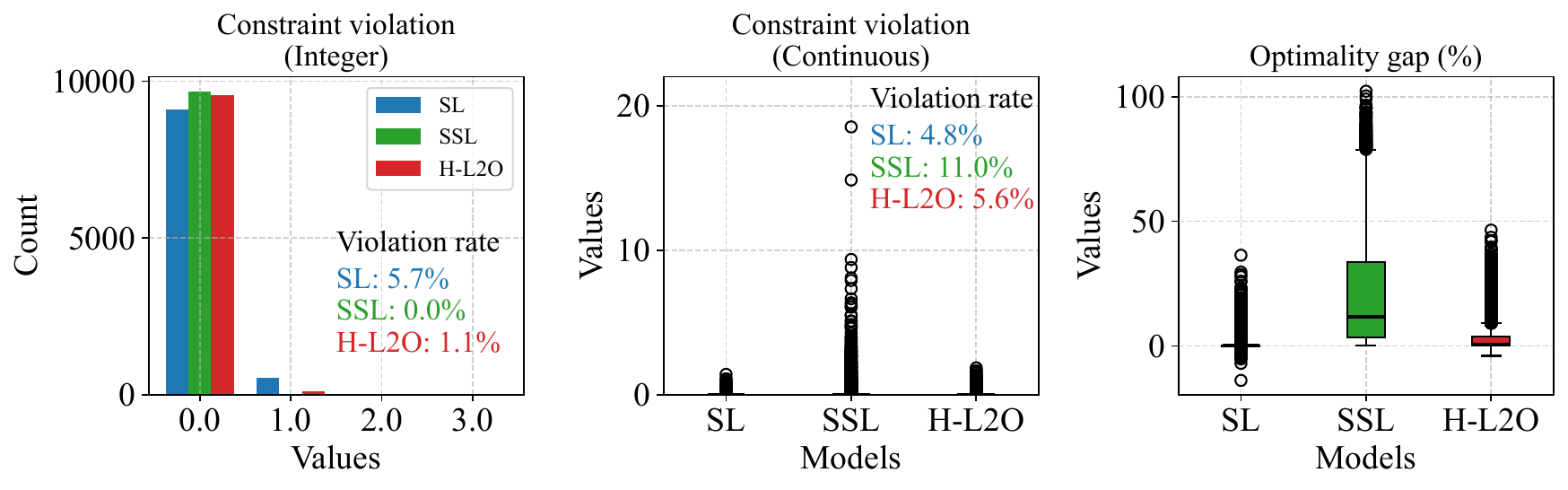}
\caption{Statistical comparison of the three models: hybrid L2O (H-L2O), supervised learning (SL), and self-supervised learning (SSL), for thermal energy tank example.}
\label{fig:thermal_tank}
\vspace{-5mm}
\end{figure}

We compare the proposed hybrid L2O framework with an SL model and an SSL model. 
Note that the SSL model considered here follows the architecture of our framework with the differentiable QP layer, rather than the conventional design in which the NN is trained to predict both integer and continuous solutions. 
However, the loss function used to train the SSL model does not include the supervised term. We evaluate the trained models using two metrics: constraint violation and optimality gap. 
We separate the violations into those associated with constraints involving continuous variables and those involving only integer variables. 
The optimality gap is expressed as a percentage, computed as the ratio of the objective gap to the optimal objective value.
We show the statistical comparison for the two examples in Figures~\ref{fig:robot_nav} and \ref{fig:thermal_tank}, respectively.
In each figure, the left panel shows the violations for integer-only constraints in the form of barplots, while the boxplots in the middle and right panels show the continuous-constraint violations and the optimality gap, respectively.
For the figures showing constraint violations, we also report the violation rate, \ie the percentage of validation problems in which the obtained solutions violate the constraints.
For the robot navigation example, the results indicate that all three models satisfy the constraints involving only integer variables. 
However, for continuous-constraint violations, the SL model fails to ensure constraint satisfaction in $71.7\%$ of the test cases (a consequence of SL training only for the integer solution, which does not guarantee a feasible continuous solution from the subsequent QP), whereas the hybrid L2O and SSL models exhibit much smaller violation rates of $8\%$ and $0.6\%$, respectively, demonstrating improved constraint satisfaction.
The plot of the optimality gap shows that the hybrid L2O model achieves better optimality than the SSL model, although it exhibits a slightly larger gap than the SL model. 
A similar trend in optimality is observed in the second example. 
Regarding constraint satisfaction, the hybrid L2O model outperforms the SL model for constraints involving only integer variables, achieving a lower violation rate of $1.1\%$ compared to $5.7\%$.
Meanwhile, the two models achieve comparable levels of satisfaction for continuous constraints ($5.6\%$ vs. $4.8\%$ violation rate). 
In contrast, the SSL model perfectly satisfies the integer-only constraints but performs poorly on the continuous ones. 
Overall, the results show that the proposed hybrid L2O framework effectively balances feasibility and optimality, achieving optimality comparable to supervised learning while improving constraint satisfaction.

Finally, we report the computation times in Table~\ref{tab:solve_time}, comparing the L2O approach (either SL, SSL, or the hybrid L2O) with the \texttt{GUROBI} solver \citep{gurobi}.
Note that the computation time for L2O approach includes both the NN prediction time and the time required to solve the relaxed QP problem.
The results confirm that the L2O approach significantly reduces the overall solving time compared to a state-of-the-art MIQP solver. 
In particular, the L2O approach achieves approximately $9 \times$ and $12 \times$ faster computation for the robot navigation and energy tank examples, respectively.

Although our proposed hybrid L2O framework demonstrates some benefits over SL and SSL approaches, it still has certain limitations that can be addressed in future research. 
First, the choice of weights in the hybrid loss function significantly affects the optimality and feasibility performance and currently requires manual tuning. 
Second, integrating the differentiable QP layer considerably increases the training time compared to purely supervised learning. 
Finally, the current framework is limited to MIQPs, while integrating differentiable optimization layers into general mixed-integer convex or nonlinear programming problems remains an open challenge.

\setlength{\textfloatsep}{10pt}   %
\begin{table}[t]
\centering
\caption{Average solving time and standard deviation for \texttt{GUROBI} solver and the L2O solver.}
\label{tab:solve_time}
\renewcommand{\arraystretch}{1.2} %
\begin{tabular}{
>{\centering\arraybackslash}p{4.5cm}  %
>{\centering\arraybackslash}p{2.5cm}  %
>{\centering\arraybackslash}p{3.0cm}  %
>{\centering\arraybackslash}p{3.0cm}  %
}
\toprule
\textbf{Problem} & \textbf{Solver} & \textbf{Avg.\ time (ms)} & \textbf{Std.\ dev. (ms)} \\
\midrule
Robot navigation example & \texttt{GUROBI}   & $69.49$ & $19.06$ \\
                         & L2O  & $7.55$  & $1.31$  \\
\midrule
Energy tank example      & \texttt{GUROBI}   & $15.40$ & $8.61$  \\ 
                         & L2O  & $1.31$  & $0.39$  \\  
\bottomrule
\end{tabular}
\end{table}

\section{Conclusions}

In this work, we developed a hybrid L2O framework for MIQPs that integrates a differentiable QP layer and a hybrid loss function combining supervised learning and self-supervised learning. We designed a model architecture in which a neural network learns the mapping from problem parameters to optimal integer variables, while the differentiable QP layer computes the corresponding continuous variables. 
This architecture enables the optimization structure to be incorporated into the learning process. 
To balance solution optimality and constraint feasibility during training, we defined a hybrid loss that linearly combines supervised and self-supervised terms. 
We validated the framework on two benchmark MPC examples, showing that the hybrid L2O approach effectively balances optimality and feasibility: it achieves better constraint satisfaction than purely supervised learning and better optimality than purely self-supervised learning. 
Future work will focus on extending the framework to mixed-integer convex and nonlinear programming problems.

\acks{This work was partially supported by US DoT Safety21 National University Transportation Center and NSF grants CISE-$2431569$.}

\bibliography{references,references_IDS}

\appendix

\section{Proof of Theorem \ref{thrm:inf}}
\label{apx:proof}

\begin{proof}
We first prove (ii) given assuming that $\bbsym{x}^*_\rho \in S_v$.
Let $S_v = \{ \bbsym{x} \,|\, v(\bbsym{x}) = v^* \}$ be the set of all points that achieve this minimum violation.
$S_v$ is closed since it is a level set of a continuous function.
Combining with the condition that either $f(\bbsym{x})$ is coercive or $\XXX$ is compact, there exist a minimizer of $f(\bbsym{x})$ on $S_v$.
From the definition of $\bbsym{x}^*_\rho$, we have
\begin{equation}
\label{eq:x_rho}
\bbsym{x}^*_\rho \in \underset{\bbsym{x} \in \XXX}{\argmin} \; f(\bbsym{x}) + \rho v(\bbsym{x}),
\end{equation}
which means $\bbsym{x}^*_\rho$ is also a minimizer of $f(\bbsym{x}) + \rho v(\bbsym{x})$ on $S_v$, \ie
\begin{equation}
\bbsym{x}^*_\rho \in \underset{\bbsym{x} \in S_v}{\argmin} \; f(\bbsym{x}) + \rho v(\bbsym{x}) = \underset{\bbsym{x} \in S_v}{\argmin} \; f(\bbsym{x}) + \rho v^*.
\end{equation}
Since $\rho v^*$ is a constant, then $\bbsym{x}^*_\rho \in \underset{\bbsym{x} \in S_v}{\argmin} \; f(\bbsym{x})$, or $\bbsym{x}^*_\rho$ is a minimizer of $f (\bbsym{x})$ on $S_v$.

For (i), we prove that for a sufficiently large but finite $\rho$, $\bbsym{x}_\rho^*$ must be in $S_v$.
Let us assume, for the sake of contradiction, $\bbsym{x}_\rho^* \notin S_v$.
From \eqref{eq:x_rho}, we have
\begin{equation}
\label{eq:rho-bound}
f(\bbsym{x}^*_\rho) + \rho v(\bbsym{x}^*_\rho) \le f(\bbsym{x}^{**}) + \rho v(\bbsym{x}^{**}) = f(\bbsym{x}^{**}) + \rho v^*,
\end{equation}
for any $\bbsym{x}^{**} \in S_v$, which leads to 
\begin{equation}
\rho \le \frac{f(\bbsym{x}^{**}) - f(\bbsym{x}^*_\rho)}{v(\bbsym{x}^*_\rho) - v^*}.
\end{equation}
Next, since $v(\cdot)$ is a convex piecewise linear function, the Hoffman error bound property holds, \ie
\begin{equation}
v(\bbsym{x}^*_\rho) - v^* \ge c\; \mathrm{dist} (\bbsym{x}^*_\rho, S_v),
\end{equation}
with $c > 0$.
Moreover, $f(\bbsym{x})$ is a quadratic function, so it is Lipschitz continuous, and we have the following inequality
\begin{equation}    
|f(\bbsym{x}^{**}) - f(\bbsym{x}_{\rho}^*)| \le L_f \cdot \|\bbsym{x}^{**} - \bbsym{x}_{\rho}^*\|. 
\end{equation}
Therefore, if we choose $\bbsym{x}^{**} \in S_v$ such that $\mathrm{dist} (\bbsym{x}^*_\rho, S_v) = \|\bbsym{x}^{**} - \bbsym{x}_{\rho}^*\|$, \ie $\bbsym{x}^{**}$ is the closest point in $S_v$ to $\bbsym{x}$, then from \eqref{eq:rho-bound}, we obtain that $\rho \le \frac{L_f}{c}$ must be satisfied.
Thus, we can select $\rho$ such that $\rho > \frac{L_f}{c}$, which leads to a contradiction.
The proof is thus complete.
\end{proof}

\section{Details of Numerical Examples}
\label{apx:example}
\subsection{Collision Avoidance for Robot Navigation}

In this example, we consider a navigation problem for a single robot operating in an environment with stationary obstacles.
The robot is required to move from its initial position to a designated goal while avoiding collisions with obstacles.
This problem is formulated as an MIQP, where binary variables are used to formulate the collision avoidance constraints between the robot and the obstacles.
The corresponding MI-MPC formulation follows the setup described in \citep{le2025multirobot}.
 
We consider an MPC problem with a single robot and $n_o$ obstacles, and let $\mathcal{O}$ be the set of obstacles.
We formulate the MPC problem with a control horizon of length $H$.
Let $t \in \ZZplus$ be the current time step.
At every time step $k \in \ZZplus$, let $\bbsym{p}(k) = [p^x(k),p^y(k)]^\top \in \RR^2$, $\bb{v}(k) = [v^x(k),v^y(k)]^\top \in \RR^2$, and $\bbsym{u}(k) = [u^x(k),u^y(k)]^\top \in \RR^2$ be the vectors of positions, velocities, and accelerations for robot, respectively.
Let $\bbsym{x}(k) = [\bbsym{p}(k),\bb{v}(k)]^\top$ be the state vector of robot.
The dynamics of each robot are governed by a discrete-time double-integrator model as follows,
\begin{equation}
\label{eq:dynamics}
\begin{split}
\bbsym{p} (k+1) &= \bbsym{p} (k) + \tau \bbsym{v} (k) + \frac{1}{2} \tau^2 \bbsym{u} (k), \\
\bbsym{v} (k+1) &= \bbsym{v} (k) + \tau \bbsym{u} (k),
\end{split}
\end{equation}
where $\tau \in \RR_{> 0}$ is the sampling time period, and compactly expressed as $\bbsym{x} (k+1) = \bbsym{f} (\bbsym{x} (k), \bbsym{u} (k))$.
We assume that the states and control inputs of robots are subjected to the following bound constraints: %
\begin{equation}
\label{eq:bound}
\begin{split}
p^x_{\min} & \le p^x (k) \le p^x_{\max},\quad
p^y_{\min} \le p^y (k) \le p^y_{\max},\\    
-v_{\max} & \le v^x (k), v^y (k) \le v_{\max},\quad  
-a_{\max} \le u^x (k), u^y (k) \le a_{\max},
\end{split}
\end{equation}
where $[p^x_{\min}, p^x_{\max}, p^y_{\min}, p^y_{\max}]^\top \in \RR^4$ is the boundary of the environment, $v_{\max} \in \RR_{> 0}$ and $a_{\max} \in \RR_{> 0}$ are the maximum speed and acceleration of the robots, respectively.
More compactly, \eqref{eq:bound} is expressed as $\bbsym{x}(k) \in \XXX$ and $\bbsym{u}(k) \in \UUU$.

The mixed-integer constraints for collision avoidance between the robot and obstacle $o \in\mathcal{O}$ at time-step $k$ are formulated by big-M formulation as follows,
\begin{equation}
\label{eq:obs_ca}
\begin{split}
\cos \alpha_o (p^x(k+1) - p^x_o) + \sin \alpha_o (p^y(k+1) - p^y_o) &\ge L_o + d_{\min} - M b_{1,o}(k), \\
- \sin \alpha_o (p^x(k+1) - p^x_o) + \cos \alpha_o (p^y(k+1) - p^y_o) &\ge W_o +  d_{\min} - M b_{2,o}(k), \\
-\cos \alpha_o (p^x(k+1) - p^x_o) - \sin \alpha_o (p^y(k+1) - p^y_o) &\ge L_o + d_{\min} - M b_{3,o}(k), \\
\sin \alpha_o (p^x(k+1) - p^x_o) - \cos \alpha_o (p^y(k+1) - p^y_o) &\ge W_o + d_{\min} - M b_{4,o}(k), \\
\end{split}
\end{equation}
where $b_{1,o}(k)$, $b_{2,o}(k)$, $b_{3,o}(k)$ and $b_{4,o}(k)$ are binary decision variables satisfying
\begin{equation}
b_{1,o}(k) + b_{2,o}(k) + b_{3,o}(k) + b_{4,o}(k) \le 3,    
\end{equation}
$[p^x_o, p^y_o]^\top$ is the center location, $\alpha_o$ is the rotation angle, and $2L_o$ and $2W_o$ are the length and width of \obstacle{o}$\in \OOO$, respectively, while $d_{\min}$ is the minimal distance between robot and obstacle to be considered as no collision.
We define the binary decision variables $\bbsym{\delta}(k) \in \{0,1\}^{4n_o}$ at each time step $k$ as the concatenated vector of $b_{1,o}(k), b_{2,o}(k), b_{3,o}(k), b_{4,o}(k)$, for all $o \in \OOO$, and rewrite all the collision avoidance constraints as $ \bbsym{g}_o(\bbsym{x}_{k+1}, \bbsym{\delta}_k) \le 0$. 

The objective for the robot is to reach the goal, \ie minimize the distance to the goal, while maintaining the minimum effort.
Thus, we consider the following MPC cost given by a weighted sum of terminal cost $\bar{c}$ and running cost $c$ over the horizon \ie
\begin{equation}
\begin{multlined}
\underset{\substack{\bbsym{x}(k+1) \in \XXX,\\ \bbsym{u} (k) \in \UUU}}{\minimize} \;\; \bar{c} \big(\bbsym{x} (t+H)\big) + \!\sum_{k=t}^{t+H-1} c \big(\bbsym{u}(k), \bbsym{x}(k) \big),
\end{multlined}
\end{equation}
where
\begin{equation}
\begin{split}
&\bar{c} (\bbsym{x}(t+H)) = \omega_{\mathrm{pt}} \left\Vert \bbsym{p}(t+H) - \bbsym{p}_{\rm{g}} \right \Vert_2^2, \\
&c (\bbsym{u}(k), \bbsym{x}(k)) = \omega_{\mathrm{p}}  \left\Vert \bbsym{p}(k) - \bbsym{p}_{\rm{g}} \right \Vert_2^2 + \omega_{\mathrm{u}} \left\Vert \bbsym{u}(k) \right\Vert_2^2
\end{split}
\end{equation}
with $\bbsym{p}_{\rm{g}}$ being the vector of goal positions, while $\omega_{\mathrm{pt}}$, $\omega_{\mathrm{p}}$, and $\omega_{\mathrm{u}}$ are positive weights.
Consequently, the cost function is quadratic in the continuous decision variables.

Therefore, the parametric MI-MPC problem can be given by
\begin{align}
\begin{split}
\text{minimize} \quad & \bar{c}(\bbsym{x}_H; \boldsymbol{\theta})
+ \sum_{k=0}^{H-1} c(\bbsym{x}_k, \bbsym{u}_k; \boldsymbol{\theta}) \\
\text{subject to} \quad
& \bbsym{x}_0 = \bbsym{x}_\text{init}(\boldsymbol{\theta}), \\
& \bbsym{x}_{k+1} = \bbsym{f}(\bbsym{x}_k, \bbsym{u}_k), \\
& \bbsym{g}_o(\bbsym{x}_{k+1}, \bbsym{\delta}_k) \le 0, \\
& \bbsym{x} \in \mathcal{X}^{H+1}, \;
\bbsym{u} \in \mathcal{U}^{H}, \;
\bbsym{\delta}_k \in \{0,1\}^{4n_o}.
\end{split}
\end{align}
where the parameter vector $\boldsymbol{\theta} = [\bbsym{x}_0^\top, \bbsym{p}_{\rm g}^\top]^\top \in \mathbb{R}^6$ contains the initial state and goal position.
We consider an MPC problem with three obstacles and a control horizon of length 20, leading to
${3 \times 20 \times 4 = 240}$ binary decision variables.
An NN $\bbsym{\pi}_{\bbsym{\omega}}: \mathbb{R}^6 \rightarrow \{0,1\}^{240}$ is employed to predict the binary decision variables.
The parameters for the simulation are set as follows:
$\tau = \SI{0.25}{s}$, 
$[p^x_{\min}, p^x_{\max}, p^y_{\min}, p^y_{\max}]^\top = [\SI{-0.5}{m}, \SI{3}{m}, \SI{-3}{m}, \SI{0.5}{m}]^\top $, 
$v_{\max} = \SI{0.5}{m/s}$ , 
$a_{\max} = \SI{0.5}{m/s^2}$, 
$d_{\min} = \SI{0.25}{m}$,
$M = 10^3$, 
$\omega_{\mathrm{pt}} = 10$, $\omega_{\bbsym{p}} = 1$, $\omega_{\mathrm{u}} = 1$, $w_s = 10^4$. 
The information of the three obstacles is: 
\begin{enumerate}
\item Obstacle 1: $p_1^x = \SI{1.0}{m}$, $p_1^y = \SI{0.0}{m}$, $L_1 = \SI{0.8}{m}$, $W_1 = \SI{1.0}{m}$, $\alpha_1 = \SI{0.0}{rad}$.
\item Obstacle 2: $p_2^x = \SI{0.7}{m}$, $p_2^y = \SI{-1.1}{m}$, $L_2 = \SI{1.0}{m}$, $W_2 = \SI{0.8}{m}$, $\alpha_2 = \SI{0.0}{rad}$.
\item Obstacle 3: $p_3^x = \SI{0.4}{m}$, $p_3^y = \SI{-2.5}{m}$, $L_3 = \SI{0.8}{m}$, $W_3 = \SI{1.0}{m}$, $\alpha_3 = \SI{0.0}{rad}$.
\end{enumerate}

\subsection{Simplified Thermal Energy Tank System}

In the second example, we examine a simplified thermal energy tank system, adapted from \citep{boldocky2025learning}. 
The system dynamics are represented by a discrete-time linear time-invariant (LTI) model with both continuous and discrete control inputs, described as follows:
\begin{equation}
\bbsym{x}_{k+1} = 
    \bbsym{A}\bbsym{x}_k 
    + \bbsym{B}_u \bbsym{u}_k 
    + \bbsym{B}_d \delta_k 
    + \bbsym{E} \bbsym{d}_k,
\end{equation}
where $\bbsym{x}_k \in \mathbb{R}^2$ is the state vector, $\bbsym{u}_k \in \mathbb{R}^2$ is the continuous control input, $\delta_k \in \{0,1,2,3\}$ is the discrete control input, and $\bbsym{d}_k \in \mathbb{R}^2$ represents known disturbances at time step $k$. 
The dynamics matrices $\bbsym{A}$, $\bbsym{B}_u$, $\bbsym{B}_d$, and $\bbsym{E}$ are:
\begin{equation}
    \bbsym{A} = 
    \begin{bmatrix} 
    0.9983 & 0.001 \\ 
    0 & 0.9966 
    \end{bmatrix}, \quad
    \bbsym{B}_u = 0.075\, \II_2, \quad
    \bbsym{B}_d = 
    \begin{bmatrix} 
    0 \\ 
    0.0825 
    \end{bmatrix}, \quad
    \bbsym{E} = -0.0833 \, \II_2.
\end{equation}
The system is subject to the following state and input constraints:
\begin{align}
\begin{split}
    & 0 \le x_{1,k} \le 8.4, \quad 
    0 \le x_{2,k} \le 3.6, \quad 
    0 \le u_{1,k}, u_{2,k} \le 8.
\end{split}
\end{align}
In addition to the constraints on the continuous control inputs, we also impose constraints on the changes between consecutive time steps of the discrete control input as follows,
\begin{equation}
-1 \le \delta_{k} - \delta_{k-1} \le 1,\; k = 1, \dots, H-1
\end{equation}
The control objective is to minimize the expected cumulative cost over the prediction horizon, which includes both state tracking and control effort penalties. The stage cost function and terminal cost function are defined as
\begin{align}
\begin{split}
\bar{c}
(\bbsym{x}_k,\bbsym{u}_k,\delta_k; \bbsym{r}_k) 
&= 
\left\|\bbsym{x}_k - \bbsym{r}_k\right\|^2_{\bbsym{Q}}
+ \left\|\bbsym{u}_k\right\|^2_{\bbsym{R}}
+ \rho \left\|\delta_k\right\|^2_{2}, \\
c (\bbsym{x}_H;\bbsym{r}_H) 
&= 
\left\|\bbsym{x}_H - \bbsym{r}_H\right\|^2_{\bbsym{Q}_t},
\end{split}
\end{align}
where $\bbsym{r}_k$ is the reference state at time step $k$, and $\bbsym{Q}$, $\bbsym{R}$, $\bbsym{Q}_t$ and $\rho$ are weighting matrices and vectors, respectively, given by
$\bbsym{Q} = \bbsym{Q}_t = \II_2$, $\bbsym{R} = 0.5\,\II_2$, and $\rho = 0.1$.
For simplicity, we assume that the reference state $\bbsym{r}_k$ remains constant over the prediction horizon, \ie $\bbsym{r}_k = \bbsym{r}$ for all $k$, where
$\bbsym{r} = [4.2, 1.8]^\top$.
Thus, the optimization problem is parameterized by the current state $\bbsym{x}_t$ and the sequence of known future disturbances over the prediction horizon, $[\bbsym{d}_k, \bbsym{d}_{k+1}, \ldots, \bbsym{d}_{k+H-1}]$. 
Accordingly, the parameter vector is defined as
$\boldsymbol{\theta} = [\bbsym{x}_k^\top, \bbsym{d}_k^\top, \bbsym{d}_{k+1}^\top, \ldots, \bbsym{d}_{k+H-1}^\top] \in \mathbb{R}^{2H + 2}$.

The parametric MI-MPC problem is given by
\begin{align}
\begin{split}
\text{minimize} \quad & 
\bar{c} (\bbsym{x}_H; \bbsym{r}_H)
+ \sum_{k=0}^{H-1} 
c (\bbsym{x}_k, \bbsym{u}_k, \delta_k; \bbsym{r}_k) \\
\text{subject to} \quad
& \bbsym{x}_{k+1} = 
    \bbsym{A}\bbsym{x}_k 
    + \bbsym{B}_u \bbsym{u}_k 
    + \bbsym{B}_d \delta_k 
    + \bbsym{E} \bbsym{d}_k, \\
& -1 \le \delta_{k} - \delta_{k-1} \le 1,\; k = 1, \dots, H-1\\ 
& \bbsym{x}_k \in \mathcal{X}, \;
\bbsym{u}_k \in \mathcal{U}, \;
\delta_k \in \{0,1,2,3\}.
\end{split}
\end{align}
An NN is trained to predict the integer control inputs, \ie
$\bbsym{\pi}_{\bbsym{\omega}}: \mathbb{R}^{2H + 2} \rightarrow \{0,1,2,3\}^{H}$.
For this example, we set the control horizon to $H = 20$, resulting in a parameter vector $\boldsymbol{\theta} \in \mathbb{R}^{42}$ and an NN output dimension of $20$.
\end{document}